# Phase liquid turbulence as novel quantum approach

Sergey Kamenshchikov

Moscow State University of M.V. Lomonosov, Physical department,
1/2, Leninskie Gory, Russia, Moscow, ZIP 119991, kamphys@gmail.com

**Abstract**

In this paper we consider a nonlinear stochastic approach to the description of quantum systems. It is shown that a possibility to derive quantum properties - spectrum quantization, zero point positive energy and uncertainty relations, exists in frame of Zaslavsky phase liquid. This liquid is considered as a projection of continuous turbulent medium into a Hilbert phase space. It has isotropic minimal diffusion defined by Planck constant. Areas of probability condensation may produce clustering centers – quasi stable particles-attractors which preserve boundaries and scale-free fractal transport properties. The stability of particles has been shown in frame of the first order perturbation theory. Quantum peculiarities of considered systems have been strictly derived from markovian Fokker-Planck equation. It turned out that the positive zero point energy has volumetric properties and grows for higher time resolutions. We have shown that a quasi stable attractor may be applied as a satisfactory model of an elementary quantum system. The conditions of attractor stability are defined on the basis of Nonlinear Prigogine Theorem. Finally the integrity of classical and quantum approaches is recovered: existence of particles is derived in terms of Zaslavsky quantum fluid.

## I. Transport model

Let us introduce a nonlinear Fokker-Planck model of transport for the description of Zaslavsky phase liquid [1] evolution. We suppose that a liquid is a projection of continuous medium into a Hilbert phase space: each particle corresponds to a phase liquid coordinate-momentum pair: $\vec{X}(t) = \left(\vec{x}(t), \vec{p}(t)\right)$. Here $\vec{X}(t)$ is a characteristic vector which defines a dynamic state of the considered system. If we analyze only one dimensional case then a given nonlinear Fokker – Planck equation can be received based on the following assumptions:

- $W(X',t'|x,t) = W(X',X,t'-t) = W(X',X,_\Delta t)$. A transitional probability doesn't depend on the initial time point;
- $P(X',t) = W(X',X,t)$. A final probability density doesn't depend on the initial coordinate;
- The initial distribution density is defined by Dirac function $W(x) = \delta(0)$: initial coordinate is defined accurately in relation to the system's typical size.

In case of active phase trajectory mixing – phase liquid turbulence, a mutual correspondence of transport properties and the characteristic vector is absent: diffusion factor $B$ is an explicit function of time parameter: $B = B(X,t)$. In this case the same transport properties are inherent to the different elements of a phase space. A Fokker-Planck transport model expresses diffusion in the following way:

$$B(X,t) = \lim_{\Delta t \to 0}\left(\frac{\langle\langle_\Delta X^2\rangle\rangle}{_\Delta t}\right) \quad \langle\langle_\Delta X^2\rangle\rangle = \int (X - X_0)^2 W(X, X_0, _\Delta t) dX_0 \qquad (1)$$

Here $X_0$ is some initial point of characteristic vector observance; integration is realized though the considered phase trajectories set. Then a nonlinear transport equation can be represented in a differential form (2).

$$\frac{\partial P(X,t)}{\partial t} = \frac{1}{2} \cdot \frac{\partial}{\partial X}\left(B(X) \cdot \frac{\partial P(X,t)}{\partial X}\right) \qquad (2)$$

It is derived in frame of markovian chain properties, expressed by Chapman – Kolmogorov relation [1]:

$$W(X_3,t_3|X_1,t_1) = \int dX_2 W(X_3,t_3|X_2,t_2) \cdot W(X_2,t_2|X_1,t_1) \qquad (3)$$

Besides basic transport properties the diffusion $B$ allows definition of averaged stochastic energy $\langle\langle \varepsilon(X,t) \rangle\rangle$. This factor expresses a displacement of characteristic vector in relation to the set of accumulated trajectories. In particular case of finite time resolution energy may be defined in the following way:

$$\langle\langle \varepsilon(X,t) \rangle\rangle = \lim_{\Delta t \to 0}\left(\frac{\langle\langle _\Delta X^2 \rangle\rangle}{_\Delta t^2}\right) = B(X,t) \cdot {_\Delta t_{\min}} \qquad (4)$$

## II. Clustering and quantization

Let's consider a uniform system: $B(X,t) = B(t)$. Then clustering properties of nonlinear Fokker-Planck transport are naturally derived in frame of Fourier space-time decomposition:

$$P_k(k,t) = \frac{1}{\sqrt{2\pi}} \int_{-\Omega}^{\Omega} \hat{P}(k,\omega) \cdot \exp(-i\omega t) d\omega \qquad (5)$$

$$P_\omega(X,\omega) = \frac{1}{\sqrt{2\pi}} \int_{-K}^{K} \hat{P}(k,\omega) \cdot \exp(-ikX) dk \qquad (6)$$

A substitution of these relations into equation (2) gives a nonlinear dispersion law: $\omega(k,t) = i \cdot B(t) \cdot (k)^2$. An allocation of real parts of [2] leads to a positive α space instability increment:

$$\text{Im}(k) = -B(t) \cdot \frac{\text{Re}(k)}{\text{Re}(\omega)} \qquad \alpha(k,t) = -\text{Im}(k) > 0 \qquad (7)$$

As a result the distribution of density tends to instable space oscillations of exponential growth:

$$P(X,t) = \int P_k(k,t) \cdot \exp(|\text{Im}(k)| \cdot X) \exp(-i\text{Re}(k) \cdot X) dk \qquad (8)$$

In terms of uniform approximation the relation (8) shows that a markovian system is converted into the set of $P(X,t)$ regular fluctuations having an exponential growth. Areas of probability condensation may be represented as clustering centers in frame of Zaslavsky phase fluid model [1]. These centers are formed by stochastic islands of elementary phase attractors. This model is defined by not stationary transport properties and nonlinear increments are certainly explicit time functions. We may consider this motion as a phase liquid turbulence while clusters as elementary vortexes.

A growth of first order clusters leads to a disturbance of diffusion distribution: uniform approximation finally becomes unacceptable. However it can still be applied to each quasi uniform cluster separately. Each cluster splits into several second order clusters: scale free clustering instability is to be continued into smaller scales. Is this process infinite or some internal clustering scale exists as viscosity scale limits the turbulence cascade of L.Richardson [3]: "Big whirls have little whirls that feed on their velocity, and little whirls have lesser whirls and so on to viscosity"? In any case we have to face a quantum resolution limit. To analyze this problem we search for a stable elementary attractor with a rigid boundary that may be considered as elementary particle. For one dimensional case the following conservative system is valid:

$$\frac{\partial P(X,t)}{\partial t} = \frac{B}{2} \cdot \frac{\partial^2 P(X,t)}{\partial X^2} \quad (9)$$

$$B = {}_\Delta t\varepsilon \int_0^L W(X,X_0,{}_\Delta t)dX_0 = const$$

$$P(0,t) = P(L,t) = const \quad X \in [0,L]$$

Mathematically this system corresponds to a well known uniform linear diffusion PDE (partial differential equation). A solution is traditionally searched in a form of the Fourier expansion, given below.

$$P(X,t) = \sum_{j=1}^{N} c_j(t) \cdot \cos\left[\left(\frac{2\pi \cdot j}{L}\right) \cdot X\right] \quad P(X,t) \geq 0 \quad (10)$$

We represent a final spectrum of diffusion factor which is given in the implicit form:

$$B_j = -\frac{2}{c_j(t)} \cdot \frac{dc_j(t)}{dt} \cdot \left(\frac{L}{\pi \cdot j}\right)^2 \quad (11)$$

Switching the small fluctuation allows us to consider the first order of a perturbation theory: $c_j(t) = c_j(0) \cdot \exp(\varphi_j \cdot t)$. A substitution of this relation into (11) gives the discrete spectrum [2]:

$$B_j = -2 \cdot \varphi_j \cdot \left(\frac{L}{\pi \cdot j}\right)^2 \quad \Delta\varepsilon_j = -\frac{\varphi_j}{\pi^3} \cdot \left(\frac{L}{j}\right)^2 \cdot \Delta\omega \quad (12)$$

Here, a circular frequency $\omega = 2\pi / {}_\Delta t_{min}$ is introduced. If we consider real values of transport then an increment $\varphi_j$ is to be negative: a particle with a rigid boundary tends to be stable and preserve a uniform diffusion. This effect provides a particle structural stability.

As the initial distribution of attractor is assumed to be uniform, then its spectral width is defined by some space-time resolution $({}_\Delta X_{min}, {}_\Delta t_{min})$:

$$_\Delta\omega_t \geq \frac{2\pi}{{}_\Delta t_{min}} \quad {}_\Delta k_X \geq \frac{2\pi}{{}_\Delta X_{min}} \quad (13)$$

A substitution of the first relation into (12) gives us the needed uncertainty limit, naturally defined by diffusion: ${}_\Delta\varepsilon_j \cdot {}_\Delta t_{min} \geq B_j$. If $B_0$ is a minimal spectral value among the set of possible modes then an uncertainty can be simplified: ${}_\Delta\varepsilon \cdot {}_\Delta t \geq B_0$. An equivalence of phase space element representation allows deriving a coordinate-momentum relation as well: ${}_\Delta p_j \cdot {}_\Delta X_{min} \geq B_0$. In such a way an elementary volume of a phase area is limited by internal transport properties of a phase liquid. What is a minimal transport factor which limits a clustering cascade? It can be evaluated in frame of the suggested perturbation model as well. Let's find a minimum of transport factor (12). The solution of a first derivative condition is represented below:

$$\partial B_j / \partial j = -2\varphi_j \cdot \left(\frac{L}{\pi j}\right)^2 \cdot \frac{1}{j} + \varphi_j' \cdot \left(\frac{L}{\pi j}\right)^2 = 0 \quad (14)$$

$$\ln(\varphi_j) = 2\ln(C \cdot j) \quad \varphi_j = C \cdot j^2$$

Here $C$ is an arbitrary negative constant, defined by probability normalization. A positive sign of a second derivative is defined by the following equation:

$$\left(\frac{L}{\pi}\right)^2 \cdot \left(\frac{C}{\varphi}\right)^3 \cdot (C - 2\varphi') \succ 0 \quad (15)$$

This equation leads to a condition of $j > 1/4$ which is automatically valid for integer values of a mode number. Finally minimal diffusion and a corresponding minimal energy of particle can be represented in a following way:

$$B_0 = \frac{2|C|L^2}{\pi^2} \quad \varepsilon_0 = \frac{2|C|L^2}{\pi^2 \, _\Delta t} \tag{16}$$

An internal diffusion limit is defined by an external scale – particle size $L$. The diffusion length $\lambda$ [m²/s] may be estimated then by the group of relations (17):

$$B_0 = \frac{2|C|L^2}{\pi^2} = \frac{\lambda^2}{_\Delta t} \quad \lambda = \frac{L}{\pi}\sqrt{2|C|\,_\Delta t} \tag{17}$$

We may notice that an internal scale is a function of descriptive parameter $_\Delta t$ and two characteristics of the dynamic system: scale of the attractor $L$ and the relaxation amplitude $|C|$ which defines rarefaction of a phase fluid (we may use a phase gas term subsequently for more comfortable visual interpretation, although a strict "fluid" concept doesn't initially assume incompressibility). The relaxation amplitude expresses a rate of phase trajectories divergence and we intuitively understand the validity of direct relation between the diffusive length and a divergence rate.

Another fundamental consequence is a minimal energy quant, corresponding to a stable particle-attractor of a phase gas. This zero point energy has volumetric properties and grows for higher time resolutions. It means that high frequency oscillations of a phase gas provide a significant contribution to the minimal energetic capacity. A positive zero point level is an evidence of internal energy of an "unfrozen" phase gas. An account of a linear dispersion law (7) shows that small scale clusters of high $j$ are responsible for this capacity. In such a way the energy resonance absorption is possible if extreme time-space resolution is achievable. A classical limit of $\omega \to 0$ and $k \to 0$ leads to a natural result: $\varepsilon_0 = 0$.

**III. Phase liquid turbulence**

Let us consider a projection of a phase gas into a visible space. All consequences, mentioned above, stay valid in a coordinate space; although a classical Hamiltonian approach can't be applied for a trajectory unambiguous description. We may introduce a direct correspondence between a stochastic uncertainty relations for a stable attractor-particle and conventional quantum relations:

$$_\Delta\varepsilon \cdot _\Delta t \geq \frac{2|C|L^2}{\pi^2} \quad _\Delta\varepsilon \cdot _\Delta t \geq \frac{\hbar}{2} \tag{18}$$

$$_\Delta p_j \cdot _\Delta X \geq \frac{2|C|L^2}{\pi^2} \quad _\Delta p_j \cdot _\Delta X \geq \frac{\hbar}{2} \tag{19}$$

The Planck factor here defines a universal value of a minimal attainable diffusion in frame of a phase gas model: $\hbar = 2B_0$. Because of quantization properties we shall refer to this concept as to the quantum phase gas. Isotropy of phase gas transport properties has been derived under the condition of rigid boundaries. It should be noted that this basic assumption is naturally satisfied for quantum measurements where the instrumental scale $L$ is present. A substitution of Planck factor allows deriving the zero point quant and a corresponding spectrum as well:

$$\varepsilon_0 = \frac{\hbar\nu}{2} \quad \varepsilon_j = -2\varphi_j\left(\frac{L}{\pi j}\right)^2 \frac{1}{_\Delta t} \tag{20}$$

Here a frequency supplementary term $\nu = 1/\,_\Delta t$ has been introduced for convenient representation of quant. An example of hydrogen like spectrum allows illustration of the key transport properties (21).

$$\varepsilon_j = -2\varphi_j \left(\frac{L}{\pi j}\right)^2 \frac{1}{_\Delta t} = Ry \frac{Z^2}{j^2} \tag{21}$$

The Rydberg constant $Ry$ and a nuclear charge $Z$ are incorporated in the relation between quantum and stochastic properties of a quasi stable atomic system. Modification of this equation allows expressing a relaxation factor in the following way:

$$\varphi_j = -\frac{Ry \cdot Z^2 \pi^2 {}_\Delta t}{2L^2} = const \tag{22}$$

Again system stability decreases if larger scales are preferred. In such a way a clustering tendency, mentioned in Section II is realized.

However we should remark the qualitative difference of two cases: a uniform stable cluster (a) and turbulent medium – "phase ocean" (b), separating these clusters. According to (1) we have the following diffusive laws correspondingly:

$$\langle\langle_\Delta X^2\rangle\rangle = 2|\varphi_j| \left(\frac{L}{\pi \cdot j}\right)^2 {}_\Delta t = B_j {}_\Delta t \quad B_j \succ B_0 \tag{23a}$$

$$\langle\langle_\Delta X^2\rangle\rangle = B(t)_\Delta t \tag{23b}$$

If we make a renormalization ${}_\Delta t \to A {}_\Delta t$ where $A = const$ then a group of the shift square can be represented in the following way:

$$D(A {}_\Delta t) = \langle\langle_\Delta X^2\rangle\rangle = D(_\Delta t)A \tag{24a}$$

$$D(A {}_\Delta t, t) = \langle\langle_\Delta X^2\rangle\rangle = D(_\Delta t, t)A \tag{24b}$$

For $t \to \infty$ and ${}_\Delta t = t - t_0$ these transport laws may be sufficiently simplified:

$$D(At) = D(t)A \tag{25a}$$

$$D(At, t) \neq D(t)A \text{ - General case} \tag{25b}$$

The first law (23a, 24a or 25a) expresses Einstein's law of a Brownian particle shift. In general case a property of self similarity is not valid for unbounded media but is always present for the stable attractor (a). According to (7) it means that a linear dispersion law is valid in the first case:

$$\text{Re}(\omega) = \frac{B_j}{\alpha} \text{Re}(k) \quad \omega = \frac{2\pi}{_\Delta t} \tag{26}$$

If we introduce a conventional term of a wave number $k = 2\pi / \Lambda$, where $\Lambda$ is a considered space scale, then (25a) equation can be modified: $D(A\Lambda) = D(\Lambda)A$. Space-time self similarity, which is a natural fractal property, is a distinctive feature of a stable clusters. This scale free property provides their stability. Complexity and collective behavior of several ranges forms attraction and integrity: "order of chaos effect" according to the concept of Klimontovich [4]. Indeed a self-similarity can be preserved only if a phase-mixing is absent on the boundary: $B \neq B(t)$, $B = const$. The absence of boundary mixing means that a probability is preserved in a certain area of a phase space and clustering occurs. A constant diffusion is to a surface of a constant energy, as it follows from (9): $\varepsilon = const$.

Energy conservancy assumes that some dissipation mechanism exists. As we consider a phase fluid model, then it is natural for us to focus on a hydrodynamic approach. The mechanism of dissipation may be considered as a generalized viscosity which is "hidden" in $B(t)$ term. Let us designate $q^+$ and $q^-$ for power input and output per system phase volume. Then energy balance condition can be formulated in the following way [5]:

$$R(t) = f\left(\vec{\Pi}(t)\right) = \frac{q^+(t)}{q^-(t)} = 1 \tag{27}$$

Here $\vec{\Pi}(t)$ is a set of control parameters. We may use an example of hydrodynamic bifurcation realized in turbulent flows. In this case all input/output energy mechanisms are provided by the flow inertial forces and by the viscous dissipation correspondingly. The basic phase parameter $R(t)$ is then a generalized case of a Reynolds number $Re$. The scheme of $\varepsilon = const$ boundary formation from instability may be represented by the set of chains, following below:

$$\uparrow q^+(t) \Rightarrow \uparrow R(t) \Rightarrow R(t) \succ 1 \Rightarrow \uparrow q^-(t) \Rightarrow \downarrow R(t) \Rightarrow R(t_1) = 1 \quad (28)$$

$$\uparrow q^-(t) \Rightarrow \downarrow R(t) \Rightarrow R(t) \prec 1 \Rightarrow \uparrow q^+(t) \Rightarrow \uparrow R(t) \Rightarrow R(t_1) = 1$$

$$\downarrow q^+(t) \Rightarrow \downarrow R(t) \Rightarrow R(t) \prec 1 \Rightarrow \downarrow q^-(t) \Rightarrow \uparrow R(t) \Rightarrow R(t_1) = 1$$

$$\downarrow q^-(t) \Rightarrow \uparrow R(t) \Rightarrow R(t) \succ 1 \Rightarrow \downarrow q^+(t) \Rightarrow \downarrow R(t) \Rightarrow R(t) \succ 1$$

Here $\uparrow$ and $\downarrow$ show finite increase and decrease of corresponding parameter. A positive feedback of input/output power is compulsory condition of bifurcation. For the case of fixed input power $q^+$ basic phase parameter stabilization can be represented in the following way:

$$R(t) = f\left(\vec{\Pi}(t)\right) = \frac{q^+(t)}{q_0^-(t_0) + q_1^-(t) + q_2^-(t) + ...} \quad (29)$$

This not stationary process represents the consequent switching of nonlinear viscous vortexes. Let us apply this model to consideration of a quantum cluster-attractor. We consider an arbitrary surface $\Omega$ in a phase space, separating areas of $R < 1$ and $R > 1$. It has been shown [5] that an attractor stability may be reduced to the condition $R \leq 1$. At the same time a balance condition $R_\Omega = 1$ is valid at the considered surface (Fig.1) where a probability flow reaches zero point and the isolation condition appears: $\Phi = -grad\left[P(X,t)\right] = 0$.

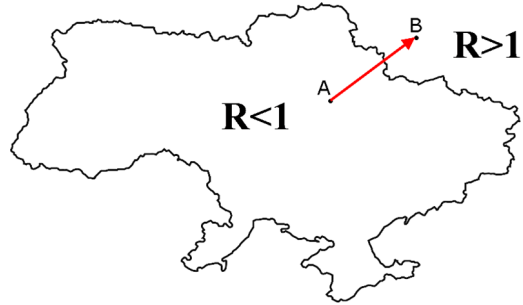

**Fig.1**. Scheme of stable attractor-cluster

A probability displacement AB is possible under the condition of $\Omega$ distortion which corresponds to a transition $R < 1 \Rightarrow R > 1$. According to Nonlinear Prigogine Theorem [5] the existence of quasi impervious boundary $\delta R > 0$ corresponds to a minimal entropy production in the vicinity of this boundary: $h \to h_{min}$. Here we introduce $h = \langle h(X(t)) \rangle$ as Kolmogorov – Sinai dynamic entropy. It is composed by averaging of positive Lyapunov factors:

$$h = \sum_{i=N}^{K} h_i^+ = \ln\left(\prod_{i=N}^{K} \sigma_i^+\right) \quad \sigma_i^+(t) = \frac{|\delta X_i(t)|}{|\delta X_i^0|} \quad (30)$$

Vector $X(t)$ is a characteristic phase vector of a system state while factor $\sigma_i^+$ shows distance growth $\delta X_i(t)$ in $i$ direction for two infinitely closely located points in phase space. In such a way it's clear that an existence of a considered surface is possible in unstable media of phase turbulence. On the other hand a surface distortion is possible only under condition of cluster instability. In Section II we have shown that cluster is stable in frame of the first order perturbation theory: cluster decay can be realized only within a higher order, strong impact.

**Conclusions**

Although we have shown that the conservancy surface may exist and is quasi stable, the nature of a phase gas viscosity is still not totally disclosed. However, it is clear that an attraction forces between particles of gas exist and contribute to appearance of viscosity and stochastic islands/attractors consequently. These compact formations, called particles-clusters, have quantum properties and internal uncertainties which make them convenient models for a quantum systems description. The charm of this approach is that a chronic collision between mechanical and quantum approaches is totally removed now. Moreover the mechanical model of a quantum phase gas gives a perspective of a global integration of Einstein and Galileo relativity principles.